\documentclass[letterpaper, 10 pt, conference]{ieeeconf}  % Comment this line out if you need a4paper

\IEEEoverridecommandlockouts                              % This command is only needed if 
\usepackage{graphicx} % for pdf, bitmapped graphics files
\graphicspath{ {./figures/} }
\usepackage{amsmath} % assumes amsmath package installed
\usepackage{amssymb}
\usepackage{algorithm}
\usepackage[noend]{algpseudocode}
\usepackage{xcolor}
\definecolor{tumblue}{RGB}{0,101,189}

\usepackage{stfloats}
\usepackage{setspace}
\usepackage{tabularx}
\usepackage{booktabs}
\usepackage[T1]{fontenc}
\usepackage[scaled=0.85]{beramono}
\usepackage{listings}
\usepackage{caption}
\usepackage{cleveref}
\usepackage{chemformula}
\usepackage{url}
\usepackage{physics}
\Crefname{lstlisting}{Listing}{Listings}
\lstdefinestyle{sparqlstyle}{
  language=OCL,
  numbers=left,
  basicstyle=\footnotesize,
  stepnumber=1,
  numbersep=10pt,
  tabsize=2,
  showspaces=false,
  breaklines=true
}
\usepackage[nonumberlist,nopostdot,acronym,toc]{glossaries}
\usepackage{soul}
\usepackage{cases}
\usepackage{balance} %insert \balance in left column of last page

\title{\LARGE \bf
A System-Level Energy-Efficient Digital Twin Framework for \\ Runtime Control of Batch Manufacturing Processes
}

\author{Hongliang Li$^{1}$, Herschel C. Pangborn$^{2}$, and Ilya Kovalenko$^{3}$% <-this % stops a space
\thanks{$^{1}$Hongliang Li is with the Department of Industrial and Manufacturing Engineering, The Pennsylvania State University, University Park, PA, USA, e-mail: hjl5377@psu.edu.}%
\thanks{$^{2}$Herschel C. Pangborn is with the Department of Mechanical Engineering, The Pennsylvania State University, University Park, PA, USA, e-mail: hcpangborn@psu.edu.}%
\thanks{$^{3}$Ilya Kovalenko is with the  Department of Mechanical Engineering and the Department of Industrial and Manufacturing Engineering, The Pennsylvania State University, University Park, PA, USA, e-mail: iqk5135@psu.edu.}%
}

\begin{document}

\maketitle
\thispagestyle{empty}
\pagestyle{empty}

%%%%%%%%%%%%%%%%%%%%%%%%%%%%%%%%%%%%%%%%%%%%%%%%%%%%%%%%%%%%%%%%%%%%%%%%%%%%%%%%
\begin{abstract}

The manufacturing sector has a substantial influence on worldwide energy consumption.
Therefore, improving manufacturing system energy efficiency is becoming increasingly important as the world strives to move toward a more resilient and sustainable energy paradigm.
Batch processes are a major contributor to energy consumption in manufacturing systems.
In batch manufacturing, a number of parts are grouped together before starting a batch process.
To improve the scheduling and control of batch manufacturing processes, we propose a system-level energy-efficient Digital Twin framework that considers Time-of-Use (TOU) energy pricing for runtime decision-making.
As part of this framework, we develop a model that combines batch manufacturing process dynamics and TOU-based energy cost.
We also provide an optimization-based decision-making algorithm that makes batch scheduling decisions during runtime.
A simulated case study showcases the benefits of the proposed framework.

\end{abstract}

%%%%%%%%%%%%%%%%%%%%%%%%%%%%%%%%%%%%%%%%%%%%%%%%%%%%%%%%%%%%%%%%%%%%%%%%%%%%%%%%
\section{Introduction}
\label{sec:introduction}

% Introduction

The manufacturing industry currently accounts for approximately one-third of worldwide energy consumption~\cite{iea2022}.
To help improve the sustainability of this sector, a number of incentives have been provided for companies to reduce the energy consumption of their manufacturing systems.
One important area of improvement for manufacturers is scheduling and control of their batch manufacturing processes~\cite{doe_incentives,eu_energy}.
Batch manufacturing processes often require long processing time and are highly energy-intensive, as they involve large and high-power equipment, such as furnaces, reactors, and mixers~\cite{FOWLER20221}.
%For example, in the semiconductor industry, batch processes are used to manufacture microchips and other electronic components~\cite{KOO20131690}.
In batch manufacturing processes, a specific quantity of products, known as a batch, is produced at one time on a batch-production machine (BPM).
The problem of optimally scheduling the batch sequence is considerably complex and has been identified as an NP-hard problem~\cite{POTTS2000228}.
Because of the energy-intensive nature of batch processes, some models and methods have been proposed that consider energy usage during batch scheduling.
For example, several scheduling methods have included energy consumption as an objective in the optimization process~\cite{9005246,9238439}.

Demand-side energy management offers another promising strategy to improve the energy efficiency of batch manufacturing processes\cite{EID201615,STRBAC20084419}.
This strategy focuses on shifting energy-intensive operations to times with a lower energy price to reduce total energy costs.
This process is often incentivized by utilities through Demand Response (DR) programs~\cite{SHARDA2021102517}.
One method of DR is Time-of-Use (TOU) energy pricing, in which the electricity price varies hourly to reflect consumer demand and the availability of renewable energy on the grid~\cite{rdab018}.
Previous works have shown that energy-efficient scheduling under TOU energy pricing can be accomplished by using model-based optimization~\cite{aghelinejad2018production, PARK2022108507,sin2020bi}. 
However, these studies solve the batch schedule problems offline and treat the corresponding schedule as fixed once production has started.
The offline scheduling does not leverage runtime information and cannot account for disturbances (e.g., change of TOU energy prices, deterioration of machines, or change of production goals), which may lead to undesired performance. The development and integration of effective runtime control strategies for batch manufacturing processes that incorporate TOU energy pricing is an ongoing research challenge.

\begin{figure}[t]
    \centering
    \includegraphics[width=0.48\textwidth]{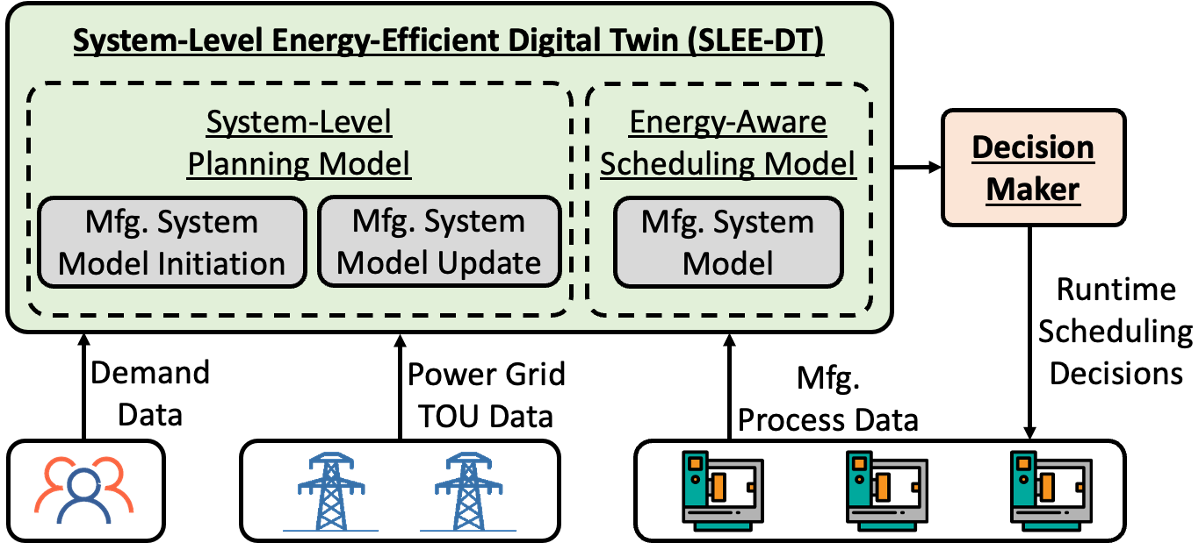}
    \caption{An overview of the proposed framework for runtime control of batch manufacturing processes.}
    \label{fig:highLevel}
    \vspace{-20pt}
\end{figure}

\begin{figure*}[!ht]
    \centering
    \includegraphics[width=0.98\textwidth]{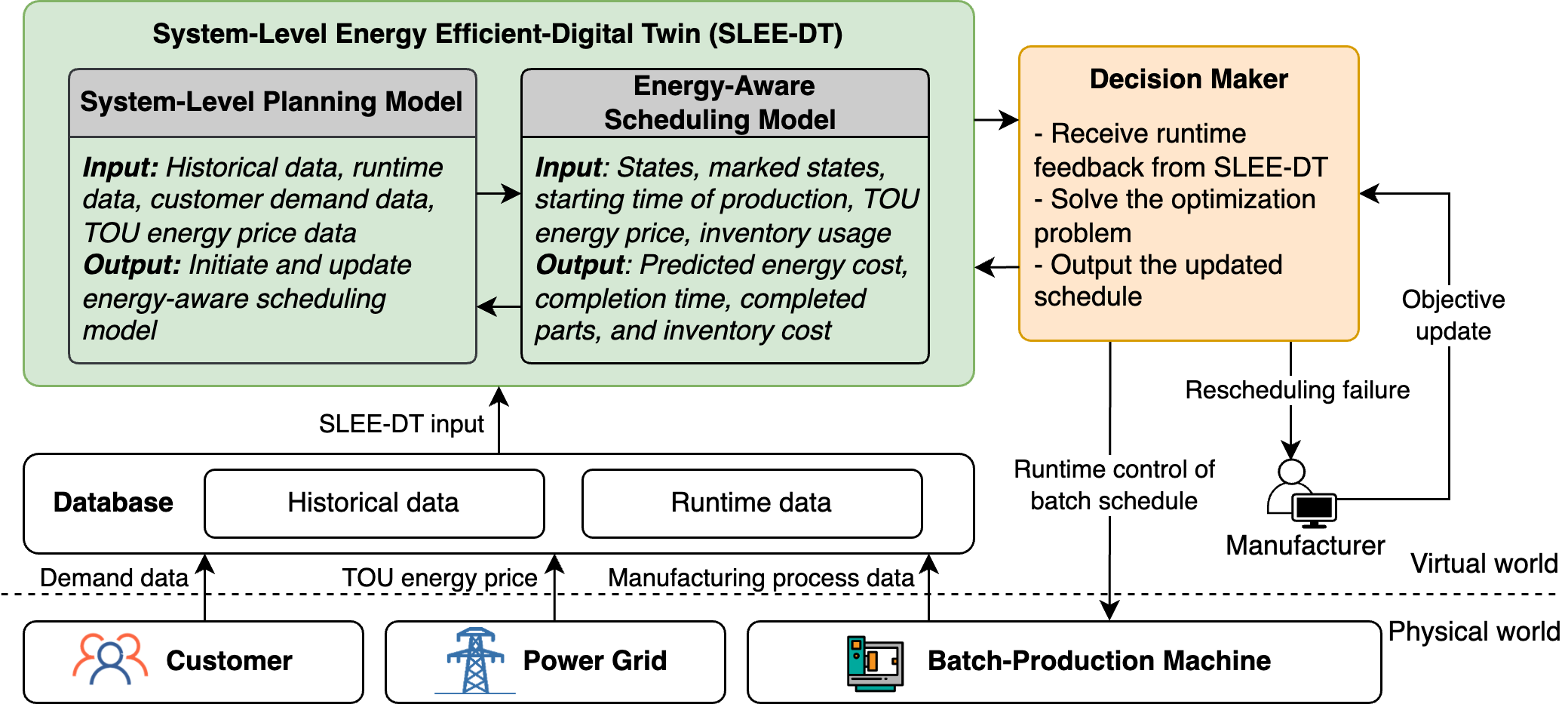} %0.98
    \caption{Overview of the SLEE-DT framework.}
    \label{fig:DTframework}
    \vspace{-15pt}
\end{figure*}

An additional challenge is the lack of models that can accurately capture and predict the dynamics of the systems and leverage runtime data for model updates. Recently, Digital Twins (DTs) have been proposed to improve real-time decision-making and control of manufacturing systems, which provides a promising solution to this challenge ~\cite{9109299,9751037}.
A DT is a purpose-driven, virtual representation of components, processes, assets, and systems that enable understanding, prediction, and optimization of performance~\cite{refId0}.
One of the highlights of DTs is the integration of modeling, simulation, and other analytical tools, which enables comprehension and prediction of manufacturing systems with greater granularity, as compared to traditional modeling methods.
Additionally, the trend of implementing DT is supported by advancements in the Internet of Things (IoT) and edge computing, which provide DTs with the capability to maintain runtime representations of manufacturing systems.
Previous studies applied DTs for specific use cases in the manufacturing industry, such as predictive maintenance~\cite{8927397} and job shop scheduling~\cite{8821409}. 

In this work, we propose a novel energy-aware runtime control strategy for batch manufacturing processes through a system-level energy-efficient DT (SLEE-DT) framework. A high-level overview of the SLEE-DT framework is shown in \Cref{fig:highLevel}.
Specifically, the major contributions of this work are (1) a discrete event system-based approach to model the batch manufacturing process and TOU-based energy costs, (2) an optimization-based decision-making model to determine the batch schedule, and (3) a system-level energy-efficient DT framework for run-time energy management and batch schedule control. The proposed framework is showcased in a simulated case study.

The remainder of this paper is as follows.
A problem statement for energy-efficient batch schedule of the batch manufacturing process is developed in \Cref{sec:problemDescription}.
\Cref{sec:dtFramework} provides a detailed description of the SLEE-DT framework.
\Cref{sec:case} presents a simulated case study to demonstrate how the proposed SLEE-DT framework can be used to improve the energy efficiency of an example batch manufacturing process.
\Cref{sec:conclusion} provides conclusions and future research directions.

\section{Problem Statement}
\label{sec:problemDescription}

% Problem formulation
This section defines the batch scheduling problem for a manufacturing system containing a BPM and a machine inventory.
Consider a BPM with one manufacturing process that can process multiple parts, e.g., a coating machine.
We define the capacity of the BPM as $H$, i.e., the machine can process at most $H$ parts at the same time. Based on this definition, let $b$ denote the batch size, where ${\{b\in \mathbb{N} : b \leq H\}}$.
We assume that there is a negligible difference in the processing time of different batch sizes for the machine, but there is a significant difference in the energy consumption rate for each size.
%The larger batch size has a higher energy consumption rate.
A higher energy consumption rate is associated with larger batch sizes.
For example, in a coating machine, there may be a set of spray nozzles that can be turned on or off depending on the number of parts inside the machine.
Every nozzle takes the same amount of time to complete the coating process, but more working nozzles require more energy usage.
We assume deterministic customer demand, i.e., the manufacturer receives a customer order to produce a certain number of products within some deadline.
The batch schedule is the sequence of batch sizes to be processed on the machine. Once a schedule is created, the unprocessed parts will be grouped into batches.
We consider a runtime batch schedule problem, which requires the manufacturer to determine a batch schedule that satisfies customer requirements with the lowest TOU-based energy costs. Note that the TOU energy prices may change during runtime.

%%%%%%%%%%%%%%%%%%%%%%%%%%%%%%%%%%%%%%%%%%%%%%%%%%%%%%%%%%%%%%%%%%%%%%%%%%%%%%%%
\section{SLEE-DT Framework}
\label{sec:dtFramework}

% DT framework
%%%%%%%%%%%%%%%%%%%%%%%%%%%%%% DT Framework figure %%%%%%%%%%%%%%%%%%%%%%%%%%%%%
An overview of the SLEE-DT framework is shown in \Cref{fig:DTframework}.
The main components of the framework are the energy-aware scheduling model, system-level planning model, decision maker, and database.

\subsection{Energy-Aware Scheduling Model}
\label{subsec:modelBPMS}
The energy-aware scheduling model encodes the batch manufacturing process dynamics and TOU-based energy costs using a priced timed automaton (PTA) model \cite{KOVALENKO2020136,9737280}.
The energy-aware scheduling model, $\mathcal{A}$, is defined as the tuple $\mathcal{A} = (Q, \Sigma, q^0, E, C, I, R, P, Q_m)$, where:
\begin{itemize}
    \item $Q=\{q^{0},q^{1},\cdots,q^{n}\}$: set of states representing the total number of parts produced by the BPM.
    \item $\Sigma = \{\sigma^{0}, \sigma^{1}, \cdots, \sigma^{H}\}$: set of events, representing a batch process of size $b$.
    \item $q^0$: initial state with $0$ parts in process.
    \item $E \subseteq Q\times \Sigma \times Q$: a finite set of transitions between states.
    \item $C$: a clock space that includes a local clock $c^l$ and global clock $c^g$.
    \item $I: Q \to \mathcal{B}(val(C))$ a mapping of states to their time-based constraints as a Boolean function of the clock valuations. 
    \item $R: E\times C \to C$: a reset operator that resets the local clock $c^l$ after each transition.
    \item $P: E \to [0, \infty)$: a mapping of a transition to its associated energy costs.
    \item $Q_m$: a set of marked states representing the amount of parts to be produced.
\end{itemize}
Note that a valuation operator, $val(\cdot)$, denotes the value of a variable, e.g., a clock or a state.

\subsubsection{Discrete Event Dynamics}
\label{subsubsec:discrete event dynamics}
$Q$, $\Sigma$, $q^0$, and $E$ represent the discrete event dynamics of the batch manuufacturing process.
A batch schedule, $s$, is defined as a sequence of events, i.e., a string over $\mathcal{A}$. We define a transition function as:
\begin{equation}
    \delta(q^c, s) \to q^f
\end{equation}
which maps the current state $q^c$ and string $s$ to a final state $q^f$ in $\mathcal{A}$.
We define two types of transitions: self-transition, and discrete transition.
$q^i \xrightarrow{\sigma^0} q^i$ is a self-transition, as  $\sigma^0$ indicates that a new batch processes $0$ part, i.e., the machine is idle, and the total number of parts, $i$, has remained the same.
$q^n \xrightarrow{\sigma^j} q^m$ is a discrete transition that indicates a change in the number of produced parts --  from $m$ parts produced to $n$ parts produced.

For example, given $\mathcal{A}$ and a schedule $s=\sigma^2\sigma^3\sigma^0\sigma^2$, consider the following transitions:
\begin{equation*}
    q^0 \xrightarrow{\sigma^2} q^2
        \xrightarrow{\sigma^3} q^5
        \xrightarrow{\sigma^0} q^5
        \xrightarrow{\sigma^2} q^7
\end{equation*}
In this example, we have the transition function ${\delta(q^0, s) \to q^7}$, which indicates that $7$ parts have been produced by this batch schedule. Note that in this transition path, $q^5 \xrightarrow{\sigma^0} q^5$ is an example of a self-transition and $q^0 \xrightarrow{\sigma^2} q^2$ is an example of a discrete transition.

\subsubsection{Time-Based Constraints}
\label{subsubsec:Time Con}
The clock space of $C$ contains a global clock and a local clock as $C=c^g \times c^l$. Both the global clock and the local clock are continuous states of the PTA that grow at a fixed rate~\cite{9737280}. The local clock, $c^l$, represents the time at a state and is set to $0$ after every transition by the reset operator, $R$. 
The time spent in a state from a self-transition is defined as set-up time.
The time spent in a state from a discrete transition is defined as processing time.
The global clock, $c^g$, continuously increases in the model and is never reset.
Global clock valuation of a state $q$ can be calculated based on the local clock valuations:
\begin{equation}
    val(c^g_q)=val(c^g_{q^\prime})+val(c^l_{q})
\end{equation}
where $q^\prime$ is the previous state.

Time-based constraints include local clock constraints and global clock constraints. Local clock constraints, $\mathcal{B}(val(c^l))$, limit the time that the system can be in a state (e.g., due to set-up time and processing time). Global clock constraints, $\mathcal{B}(val(c^g))$, encode the time-based demand requirements (e.g., ``order should be filled before a deadline''). We use the mapping $I$ to store the time constraints of each state.
For example, the Boolean function ${c^g_{q^4}}\leq 10$ evaluates to \textit{true} if the global clock at the state $q^4$ is less than or equal to $10$ time units. This is equivalent to a customer requirement that $4$ parts are needed before 10 time units.

\subsubsection{Demand Quantity-Based Constraints}
\label{subsubsec:Demand Con}
We assume that any products that remain after completing an order are stored in the machine's inventory.
If more parts are produced than required by an order, then the demand will still be met.
Therefore, any state that has more than the required amount of parts produced is contained in the set of marked states denoted as $Q_m$.
Let $d \in \mathbb{N}$ and $r\in \mathbb{N}$ denote the required demand quantity and the maximum capacity of the machine inventory, respectively.
Let $v$ denote the allocated capacity level of machine inventory and we have $\{v\in \mathbb{N} : v \leq r\}$.
Then the marked states can be defined as ${Q_m \subseteq Q=\{q^d, q^{d+1},\dots, q^{d+v}\}}$.

Let $\Sigma^*$ denote all possible strings over $\Sigma$ including the empty string $\varepsilon$.
The language (set of strings) generated by $\mathcal{A}$ is denoted as:
\begin{equation}
    \mathcal{L}(\mathcal{A})=\{s\in\Sigma^*: \delta(q^0,s) \in Q\}
\end{equation}
$\mathcal{L}(\mathcal{A})$ represents all possible batch schedules of the machine.
The language marked by $\mathcal{A}$ is:
\begin{equation}
    \mathcal{L}_m(\mathcal{A})=\{s\in\mathcal{L}(\mathcal{A}): \delta(q^0,s) \in Q_m\}
\end{equation}
$\mathcal{L}_m(\mathcal{A})$ represents all schedules that meet the demand quantity requirement. If a string representing a batch schedule, $s_g$, is part of the marked language $s_g \in \mathcal{L}_m$, then that batch schedule produces the number of parts required by the order.

\subsubsection{TOU-Based Energy Cost}
\label{subsubsec:TOU cost}
The energy cost of a batch schedule (string) is determined by adding up the energy cost of each individual batch (event), taking into account both the discrete event dynamics and the time series data.
The clock space of the PTA ensures the connection between these two aspects.
To calculate the TOU-based energy cost during runtime, the energy-aware scheduling model, $\mathcal{A}$, needs to be synchronized to the time of the power grid.

We use $T$ to denote the time of the power grid.
We denote $|s|$ as the length of string $s$, which is the number of events in $s$. Let $s_j \in \Sigma$ denote the $j$th event in $s$.
For a given $\mathcal{A}$ and a string $s_{[0,j]}$, a transition path is determined by $s_{[0,j]}$ based on $\mathcal{A}$, resulting in the final state of $q^f= \delta(q^0,s_{[0,j]})$.
Let the starting time of a batch schedule $s_{[0,j]}$ be the power grid time $T_0$.
We derive the starting time $T_s$ and end time $T_e$ of the transition with $i$th event $s_i$ in string $s_{[0,j]}$.
We call $s_{[0,i]}$ the prefix string of $s_{[0,j]}$.
Similarly, we call $s_{[i,j]}$ the suffix string of $s_{[0,j]}$. 
Prefix string $s_{[0,i]}$ leads to the transitions of $\mathcal{A}$ and ends at state $q^{e}$, which can be calculated using transition function as:
\begin{equation}
    q^{e}=\delta(q^0,s_{[0,i]})
\end{equation}
Similarly, prefix string $s_{[0,i-1]}$ leads to the state just before the state $q^{e}$ and can be calculated as:
\begin{equation}
    q^{e-1}=\delta(q^0,s_{[0,i-1]})
\end{equation}
The $i$th event in $s_{[0,j]}$, i.e., $s_i$ enables the transition:
\begin{equation*}
    q^{e-1} \xrightarrow{s_i} q^{e}
\end{equation*}
$T_e$, the end time of this transition, can be calculated as:
\begin{equation}
    T_e=T_0+val(c^g_{q^{e}})
\end{equation}
$T_s$, the starting time of this transition, can be calculated as:
\begin{equation}
    T_s = 
    \begin{cases}
        T_0, & \text{if $e=0$}\\
        T_s=T_0+val(c^g_{q^{e-1}}), & \text{if $e \geq 1$}
    \end{cases}
\end{equation}
Let $pr(s_i) \in \mathbb{R}_{\geq 0}$ denote the power of event $s_i$, i.e., the energy consumption rate of the batch size.
TOU energy price for the transition period can be obtained from the power grid, which is a function of the power grid time $T$ as $f(T)$.
Then the energy cost $P$ for the transition ${E(s_i):q^{e-1} \xrightarrow{s_i} q^{e}}$ can be calculated as:
\begin{equation}
    P_{E(s_i)}=\int_{T_s}^{T_e}pr(s_i)f(T)dT
\end{equation}
The total energy cost of $s_{[0,j]}$ is:
\begin{equation}
    \label{eq:total price}
    TP_{s_{[0,j]}}=\sum_{k=0}^{j}P_{E(s_k)}
\end{equation}

%%%%%%%%%%%%%%%%%%%%%%%%%%%%%System-level planning model%%%%%%%%%%%%%%%%%%%%%%%%%%%%
\subsection{System-Level Planning Model}
\label{subsec:systemplanning}
% The system-level planning model is a data-driven model responsible for initializing and updating the input parameters of the energy-aware scheduling model.
% These parameters include states, marked states, production start time, time-based constraints, demand quantity-based constraints, and TOU energy prices.
% The system-level planning model initiates the energy-aware scheduling model upon receiving customer order information.
% The system-level planning model is a data-driven model that initiates and updates the energy-aware scheduling model input including states, marked states, starting time of production, time-based constraints, demand quantity-based constraints, and TOU energy price. The system-level planning model initiates the energy-aware scheduling model when it receives customer order information.
The system-level planning model is a data-driven model responsible for initializing and updating the energy-aware scheduling model by analyzing both historical data and runtime data.
% When the batch manufacturing process is operating at the initial batch schedule, the system-level planning model keeps monitoring the system's performance. When a change in the system, for example, TOU energy price, is detected, the system-level planning model updates the energy-aware scheduling model to ensure the accurate representation of the batch manufacturing process.
During the operation of the batch manufacturing process based on the initial batch schedule, the system-level planning model continuously monitors the performance of the system. If a change in the system occurs, such as a variation in the TOU energy price, the system-level planning model promptly updates the energy-aware scheduling model. This ensures that the batch manufacturing process is accurately represented during runtime.

%%%%%%%%%%%%%%%%%%%%%%%%%%%%%%%Decision maker%%%%%%%%%%%%%%%%%%%%%%%%%%%%%%%%%%%
\subsection{Decision Maker}
\label{subsec:decisionmaker}
\begin{figure}[t]
    \centering
    \includegraphics[width=.48\textwidth]{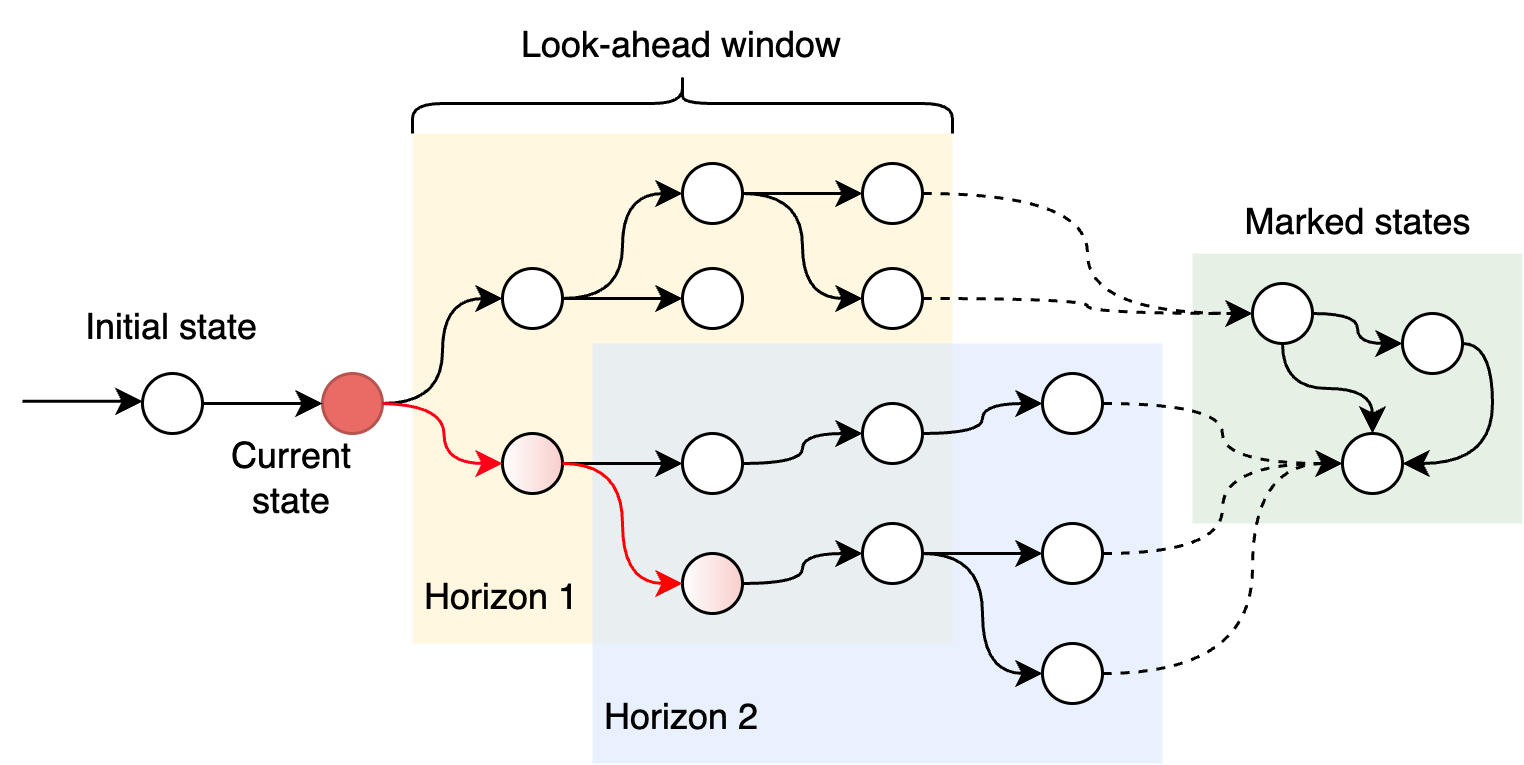}
    \caption{LLP with a 3-step look-ahead window.}
    \label{fig:look ahead}
    \vspace{-15pt}
\end{figure}

The decision maker solves an optimization problem to determine the optimal batch schedule during runtime.
We leverage a limited look-ahead policy (LLP) based online control strategy to realize the runtime decision-making.
The main idea of the LLP is illustrated in \Cref{fig:look ahead} with a $3$-step look-ahead window.
Instead of optimizing the path that minimizes the global cost, the LLP evaluates the local cost within the limited look-ahead window.
The path with the minimum cost is identified by solving an optimization problem.
Only the first event of the path is applied as the control action.
Then the window slides to the next state after the control action occurs.
Such a process is repeated until exhausting the set of marked states.
The LLP-based control strategy is a Receding Horizon Control or Model Predictive Control with a limited control horizon.
The details of the LLP can be found in \cite{182478}.
We first define the open-loop optimal control problem for the energy-aware scheduling model. Then we define the receding horizon optimization problem for the LLP at each limited look-ahead window.

\begin{figure*}[!h]
    \centering
    \includegraphics[width=0.98\textwidth]{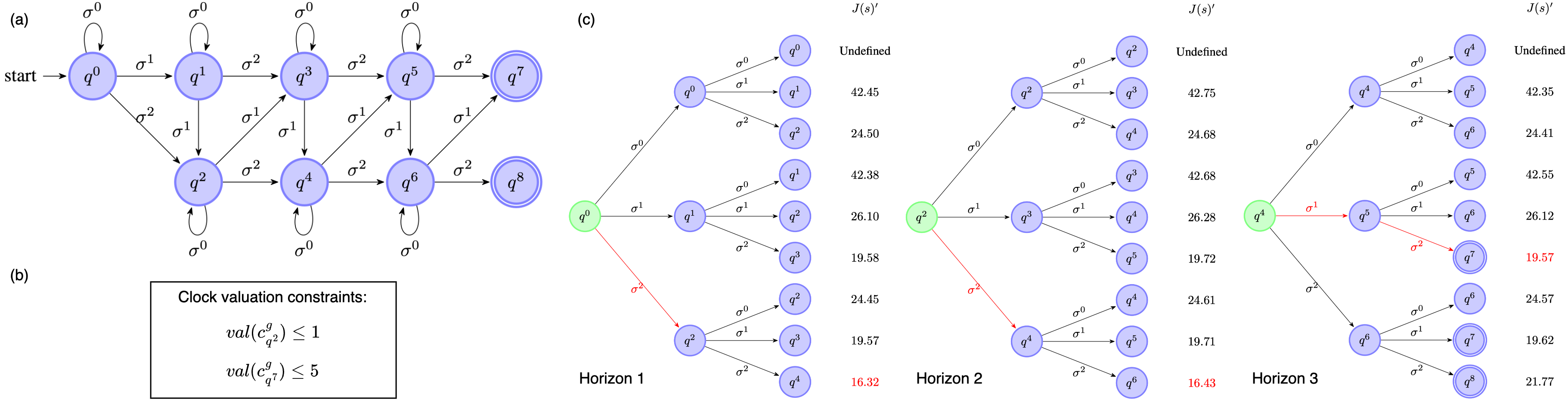} %0.98
    \caption{SLEE-DT framework for runtime control of the batch manufacturing process: (a) Energy-aware scheduling model graph used for the case study; (b) Time-based constraints; (c) LLP-based decision maker.}
    \label{fig:case-study-results}
    \vspace{-15pt}
\end{figure*}

\subsubsection{Optimal Control Problem for Energy-Aware Scheduling Model}
\label{subsubsec:treeExplore}
One of the control objectives for the batch manufacturing process modeled by PTA is to find a batch sequence with the minimum TOU-based energy cost.
%while satisfying both time-based and demand quantity-based constraints.
We employ the average energy cost as a component of the cost function, which tends to push the BPM to run at a larger batch size.
%We use average energy cost to form the cost function,
This is beneficial to maintain high machine utilization.
In \Cref{subsec:modelBPMS}, we allow the batch schedule to produce a number of parts that is greater than the demand. The remaining parts after satisfying the customer demand are stored in the machine inventory. We penalize the used capacity level of the inventory in the cost function. 
Hence, the cost function of a batch schedule $J(s)$ is defined as a two-part cost:
\begin{equation}
    J(s)=\underbrace{TP_s/{d}}_{\text{average energy cost}}+\quad \underbrace{val(\delta(q^0,s))-d}_{\text{inventory cost}} 
\end{equation}
where $TP_s$ can be calculated based on \Cref{eq:total price}, $d$ is the demand quantity.

The time-based and demand quantity-based constraints are described in \Cref{subsec:modelBPMS}. The optimal control problem is formed as:
\begin{subequations}
\begin{align}
    \label{eq:op1costf}
    \operatorname*{argmin}_{s} \quad & J(s)\\
    \label{eq:op1cons1}
    \operatorname*{s.t.} \quad & s \in \mathcal{L}_m(\mathcal{A})\\
    %& val(c^{g}_{\delta(q_0,s)}) \leq \operatorname*{Due\;date}\\
     \label{eq:op1cons2}
    & val(C) \models I
\end{align}
\end{subequations}
where $\mathcal{L}_{m}(\mathcal{A})$ is the set of accepted paths on $\mathcal{A}$. \Cref{eq:op1cons1} imposes demand quantity-based constraints while \Cref{eq:op1cons2}  imposes time-based constraints, where $\models$ indicates that the condition on the left side satisfies the condition on the right side.

\subsubsection{Limited Look-Ahead Control Strategy}
\label{subsubsec:treeExplore}
The first step of formulating the LLP is constructing a limited-step exploration of $\mathcal{A}$. Let $W$ denote the length of the look-ahead window. Let the $\Sigma^{\leq W}$ denote all the strings with a length less than or equal to $W$ as:
\begin{equation}
    \Sigma^{\leq W}=\{s\in\Sigma^*: |s|\leq W\}
\end{equation}
The strings defined in $\mathcal{A}$ and started from the current state $q^c \in Q$ is a set or a sublanguage of $\mathcal{A}$:
\begin{equation}
    \mathcal{L}_{sub}(\mathcal{A},q^c)=\{s\in \Sigma^*: \delta(q^c,s)\in Q\}
\end{equation}
The set of strings starting from the $q^c$ with length less than or equal to $W$ is defined as a $W$-step look-ahead tree $Tree(q^c,W)$:
\begin{equation}
    Tree(q^c,W)=\mathcal{L}_{sub}(\mathcal{A},q^c) \cap  \Sigma^{\leq W} 
\end{equation}
$Tree(q^c, W)$ includes all candidate strings (batch schedules) from the current state $q^c$ with $W$-step look-ahead.
The LLP performs optimizations within the limited look-ahead window to find the path with minimum cost.
When the terminal state of a string in $Tree(q^c, W)$ reaches $Q_m$, the inventory cost is included.
The cost function $J(s)^\prime$ within a $W$-step look-ahead window is formed as:
\begin{equation}
        J(s)^\prime= \left( \frac{TP_s}{(1-\xi)val(\delta(q^c,s))+\xi d}\right)
                     + \xi(val(\delta(q^c,s))-d)
\end{equation}
% \begin{equation}
%     \begin{split}
%         J(s)^\prime= &\frac{TP_s}{(1-\xi)(val(\delta(q^c,s)))+\xi d}\\
%                      &+ \xi(val(\delta(q^c,s))-d)\\
%     \end{split}
% \end{equation}
where $\xi$ is the indicator function:
\begin{equation}
    \xi = 
    \begin{cases}
        1, & \text{if $\delta(q^c,s) \in Q_m$}\\
        0, & \text{else}
    \end{cases}
\end{equation}
%Then the optimization problem for each $W$-step look-ahead tree is:
Then the optimization problem for $Tree(q^c, W)$ is:
\begin{subequations}
\label{eq:op2}
\begin{align}
    \operatorname*{argmin}_{s} \quad&
    J(s)^\prime\\
    \operatorname*{s.t.} \quad & s \in Tree(q^c, W)\\
    & val(C) \models I
\end{align}
\end{subequations}

% \begin{subequations}
% \label{eq:op2}
% \begin{align}
%     \operatorname*{argmin}_{s} \quad&
%     J(s)^\prime= \frac{TP_s}{(1-\xi)(val(\delta(q^c,s)))+\xi d}\\
%     &+ \xi(val(\delta(q^c,s))-d)\\
%     \operatorname*{s.t.} \quad & s \in Tree(q^c, W)\\
%     & val(C) \models I
% \end{align}
% \end{subequations}
% where $\xi$ is the indicator function:
% \begin{equation}
%     \xi = 
%     \begin{cases}
%         1, & \text{if $\delta(q^c,s) \in Q_m$}\\
%         0, & \text{else}
%     \end{cases}
% \end{equation}
Look-ahead exploration continues until exhaustively exploring $Q_m$.
The LLP-based online scheduling strategy may fail to generate a valid schedule, which is defined as a rescheduling failure.
When a rescheduling failure happens, the decision maker will negotiate with the manufacturer to update the production requirements or generate the schedule based on expert knowledge.

%%%%%%%%%%%%%%%%%%%%%%%%%%%%%%%%%%%%%%Database%%%%%%%%%%%%%%%%%%%%%%%%%%%%%%%%%%%%%%%%%%%%%%
\subsection{Database}
\label{subsec:database}
The database functions as a storage facility for both historical data and runtime data.
In terms of historical data, it stores previous customer order information, manufacturing process data associated with those orders, and a record of historical TOU energy prices.
The database also incorporates runtime data, which includes runtime information about the batch manufacturing process.
For example, data about processes, materials, and machines can be acquired from the Manufacturing Execution System (MES).
By housing this data, the database offers essential support to the system-level planning model enabling the initiation and continuous updating of the energy-aware scheduling model.

%%%%%%%%%%%%%%%%%%%%%%%%%%%%%%%%%%%%%%%%%%%%%%%%%%%%%%%%%%%%%%%%%%%%%%%%%%%%%%%%
\section{Case Study}
\label{sec:case}

% Case study
%The proposed framework has been applied to a case study of a simulated batch manufacturing process.
\subsection{Case Study Setup}
\label{subsec:casesetup}
Consider a manufacturing system with a BPM and a machine inventory.
The capacity of the BPM and the inventory are $H=2$ parts and $r=3$ parts, respectively.
A customer order is received that requires $2$ parts to be produced in the next $1$ hour and $5$ more parts to be produced in the following $4$ hours, i.e., a total of $7$ parts in $5$ hours.
The batch processing time is $1$ hour and the set-up time is $0.2$ hours.
The power consumption of the machine is dependent on the batch size.
For batch sizes 0, 1, and 2, these are $0.5$ megawatts (MW), $0.8$ MW, and $1$ MW, respectively.
TOU energy prices are acquired from the U.S. national grid website~\cite{NationalGrid}.
Two batch scheduling strategies are compared: the proposed method and a benchmark batch schedule based on maximizing BPM utilization without energy considerations.

\subsection{Application of the SLEE-DT Framework}
\label{subsec:results}
Once the SLEE-DT receives the order, the system-level planning model starts to analyze the system based on customer requirements and historical data.
Then, the system-level planning model initiates the energy-aware scheduling model,
$\mathcal{A}_{case}$,
with $\Sigma = \{\sigma^0,\sigma^1,\sigma^2\}$ and states $Q = \{q^0, q^1,...,q^8\}$.
The system is allowed to only use the $v=1$ part of the machine inventory, i.e., $Q_m = \{q^7, q^8\}$.
The production starts at $8$ am and the limited look-ahead window is $2$.
The time-based constraints are also generated by analyzing the order requirements.
$\mathcal{A}_{case}$ is then created as shown in \Cref{fig:case-study-results} (a) and (b).

\begin{figure}[t]
    \centering
    \includegraphics[width=.48\textwidth]{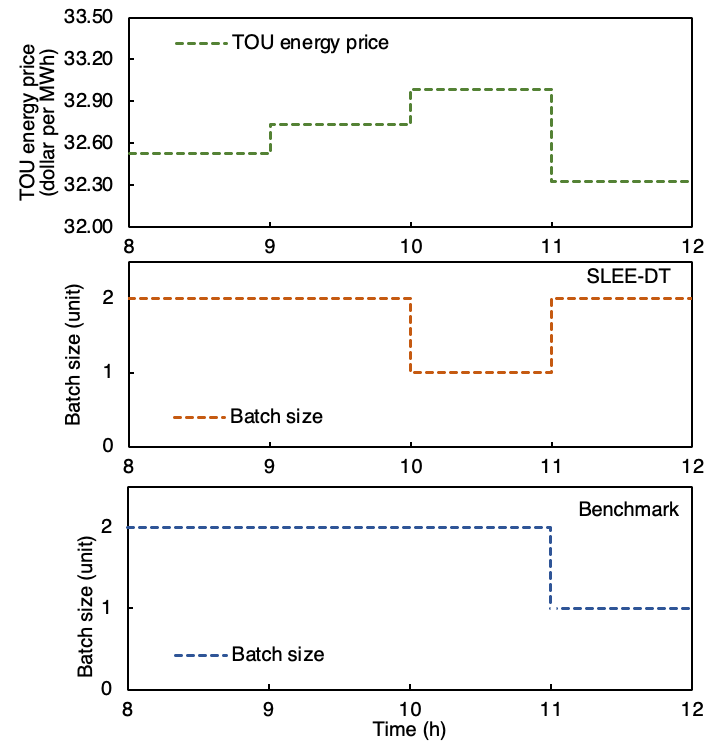}
    \caption{Batch schedule by SLEE-DT and benchmark.}
    \label{fig:caseresult}
    \vspace{-15pt}
\end{figure}

The decision maker generates the batch schedule during runtime.
As shown in \Cref{fig:case-study-results}(c), the green node is the starting state of each look-ahead window.
$J(s)^\prime$ is the cost of each path.
The cost of a path with two successive $0$ batch sizes is undefined, so the path is invalid.
The optimization problem shown in \Cref{eq:op2} is solved at each look-ahead window by exhaustive search.
The first event of the path with minimum cost, the red event as shown in \Cref{fig:case-study-results}(c), is set as the next batch size.
Note that state $q^2$ is encoded with the global clock constraint of $val(c^g_{q^2}) \leq 1$.
Therefore the transition $q^0\to q^2$ is the only valid transition. 
%Based on the LLP approach, the runtime schedule is generated and controls the batch size.
The SLEE-DT-based schedule can be represented as $\sigma^2 \to \sigma^2 \to \sigma^1 \to \sigma^2$, which is shown in \Cref{fig:caseresult}.
A batch schedule based on the benchmark strategy is also shown in \Cref{fig:caseresult}.
The energy cost of the SLEE-DT runtime schedule is 2.55\% lower than the benchmark strategy, as this shifts the energy-intensive operation of a batch size of $2$ to time when the TOU energy price is lower.

\section{Conclusion}
\label{sec:conclusion}

In this work, we propose a system-level energy-efficient Digital Twin (SLEE-DT) framework that can be used to improve the runtime control of batch manufacturing processes. 
As part of the framework, a priced timed automaton (PTA) model is used to capture both the batch process dynamics and the energy costs of the manufacturing system. An optimization-based decision maker is then proposed to enable runtime scheduling and control of batch manufacturing processes with consideration for time-of-use (TOU) energy prices.
The SLEE-DT framework reduces energy costs and improves sustainability by reallocating energy-intensive batch production to times with lower energy prices.
Future work will focus on extending the proposed framework to a larger system with multiple machines and inventories.
%integrating real-time TOU pricing data.
We will also look to implement the proposed framework in a physical manufacturing system.
%To ensure effective runtime control of the physical system, we expect to encode stochastic disturbances in the system model to capture variances between machine operations.


%%%%%%%%%%%%%%%%%%%%%%%%%%%%%%%%%%%%%%%%%%%%%%%%%%%%%%%%%%%%%%%%%%%%%%%%%%%%%%%%

%%%%%%%%%%%%%%%%%%%%%%%%%%%%%%%%%%%%%%%%%%%%%%%%%%%%%%%%%%%%%%%%%%%%%%%%%%%%%%%%



% \section*{ACKNOWLEDGMENT}

% The preferred spelling of the word ÒacknowledgmentÓ in America is without an ÒeÓ after the ÒgÓ. Avoid the stilted expression, ÒOne of us (R. B. G.) thanks . . .Ó  Instead, try ÒR. B. G. thanksÓ. Put sponsor acknowledgments in the unnumbered footnote on the first page.

\balance
%%%%%%%%%%%%%%%%%%%%%%%%%%%%%%%%%%%%%%%%%%%%%%%%%%%%%%%%%%%%%%%%%%%%%%%%%%%%%%%%
\bibliographystyle{IEEEtran}
%\begin{thebibliography}{1}

\bibliography{root}

\end{document}